\documentclass[useAMS,usenatbib]{mn2e}
\usepackage{graphicx}
\usepackage{color}
\usepackage{ulem}
\def\apj{ApJ}
\def\apjs{ApJS}

\def\aap{A\&A}
\def\aaps{A\&AS}

\def\mnras{MNRAS}

\def\pasa{PASA}

\def\pasp{PASP}
\def\nat{Nature}

\def\araa{Ann.Rev.Astron.Astrophys.}

\def\rmxaa{Revista Mexicana de Astronomia y Astrofisica}

\title[J1234$+$3901, an extremely metal-poor galaxy]
{J1234$+$3901: an extremely metal-deficient compact 
star-forming dwarf galaxy at redshift
0.133}
\author[Y. I. Izotov et al.]{Y. I.\ Izotov$^{1}$,
T. X.\ Thuan$^{2}$ and N. G.\ Guseva$^{1}$\\
                $^{1}$Bogolyubov Institute for Theoretical Physics,
                     National Academy of Sciences of Ukraine,
                     14-b Metrolohichna str., Kyiv, 03143, Ukraine,\\
                     yizotov@bitp.kiev.ua, nguseva@bitp.kiev.ua\\
                $^{2}$Astronomy Department, University of Virginia, 
                     P.O. Box 400325, Charlottesville, VA 22904-4325,\\
                     txt@virginia.edu\\
}
\begin{document}


\pagerange{\pageref{firstpage}--\pageref{lastpage}} \pubyear{2012}

\maketitle

\label{firstpage}

\begin{abstract}
We have obtained optical spectroscopy of one of the most 
metal-poor dwarf star-forming galaxies (SFG) in the local Universe, 
J1234$+$3901, with the Large Binocular 
Telescope (LBT)/Multi-Object Dual Spectrograph (MODS). This blue compact dwarf 
(BCD) galaxy with a redshift $z$=0.133 was selected from the
Data Release 14 (DR14) of the Sloan Digital Sky Survey (SDSS). Its properties 
are extreme in many ways. Its oxygen abundance   
12 + log O/H = 7.035 $\pm$ 0.026 is among the lowest ever observed for a SFG. 
Its absolute magnitude $M_g$ = $-$17.35 mag makes it the brightest galaxy 
among the known BCDs with 12 + log O/H $\la$ 7.3. 
With its low metallicity, low stellar mass $M_\star$ = 10$^{7.13}$ M$_\odot$
and very low mass-to-light ratio $M_\star$/$L_g$ $\sim$ 0.01 (in solar units), it
deviates strongly from the mass-metallicity and luminosity-metallicity relations
defined by the bulk of the SFGs in SDSS DR14. 
J1234$+$3901 has a very high  specific 
star-formation rate sSFR $\sim$ 100 Gyr$^{-1}$, indicating very active ongoing 
star-formation. Its spectrum shows a strong He~{\sc ii} $\lambda$4686 
emission line, with a flux
$\sim$ 2.4 per cent that of the H$\beta$ emission line. The most probable source 
of ionizing radiation for producing such a strong line is fast radiative shocks.
J1234$+$3901 has a ratio O$_{32}$ = [O~{\sc iii}]5007/[O~{\sc ii}]3727 $\sim$ 
15, the highest among the lowest-metallicity SFGs, and is thus likely  
leaking Lyman continuum radiation. It is a good candidate
for being a young dwarf galaxy, with a large fraction of its stars formed 
recently. As such, it is probably one of the best local counterparts of dwarf 
primeval galaxies responsible for the reionization of the early Universe.
\end{abstract}

\begin{keywords}
galaxies: dwarf -- galaxies: starburst -- galaxies: ISM -- galaxies: abundances.
\end{keywords}

\section{Introduction}\label{sec:INT}

The most metal-deficient nearby  
star-forming galaxies (SFGs), with oxygen abundances 12 + logO/H $\la$ 7.3,
are often considered to be the best local counterparts
of the dwarf galaxies at high redshifts. They share many of the same 
properties. Besides their very low metallicities, they have low stellar masses, 
high specific star formation rates, high gas masses and a compact structure. 
Therefore, these very rare nearby extremely metal-poor galaxies constitute 
excellent laboratories in 
which to study the physical conditions that prevailed in galaxies 
at redshifts $z$ $\sim$ 5--10, during the epoch of reionization of the Universe.
Because of their proximity, they can be studied with an accuracy  
that is not possible for high-redshift dwarf galaxies. 
The majority of these nearby galaxies can often
be classified as blue compact dwarf (BCD) galaxies because of their low 
mass, blue colour, due to the active star-formation, and compact morphology 
\citep{TM81}.

The first spectroscopically observed extremely metal-deficient SFG, I~Zw~18 
\citep{SS72}, has 12~+~log~O/H~$\sim$~7.17 -- 7.26 
\citep[e.g. ][]{SK93,IT98}. More recently, other SFGs with lower 
metallicities, e.g. SBS~0335$-$052W with 
12~+~logO/H = 7.12$\pm$0.03 \citep{I05} and Little Cub with 
12~+~logO/H = 7.13$\pm$0.08 \citep{H17} have been found.
The large spectroscopic data base provided by the Sloan Digital Sky Survey 
(SDSS) has allowed to increase considerably the number of known SFGs with 
12~+~log~O/H $\sim$ 7.1 -- 7.3,  to $\sim$ 50 objects
\citep[e.g. ][]{I12a,G17}. However, only very few SFGs with 
12~+~log~O/H $<$ 7.1 have been uncovered so far. One example is AGC~198691 with 
12~+~logO/H = 7.02$\pm$0.03 \citep{H16}.
Recently, \citet{I18a} have discovered in the SDSS Data Release 13 the SFG 
J0811$+$4730 with a record low 
12~+~log~O/H = 6.98$\pm$0.02, as derived from a high signal-to-noise ratio 
spectrum obtained with the Large Binocular Telescope (LBT). 
Similar or lower oxygen abundances of 7.01$\pm$0.07, 6.98$\pm$0.06, 
6.86$\pm$0.14 have been reported by \citet{I09} in three individual H~{\sc ii} regions of
SBS~0335$-$052W, and of 6.96$\pm$0.09 by \citet{An19} in one of 
the H~{\sc ii} regions in DDO~68. However, we note that the luminosity-weighted
oxygen abundances (i.e. derived from the integrated spectrum of all
observed H~{\sc ii} regions in the galaxy) in SBS~0335$-$052W and DDO~68 are 
higher than that in J0811$+$4730.

\begin{figure*}
\hbox{
\includegraphics[angle=-90,width=0.48\linewidth]{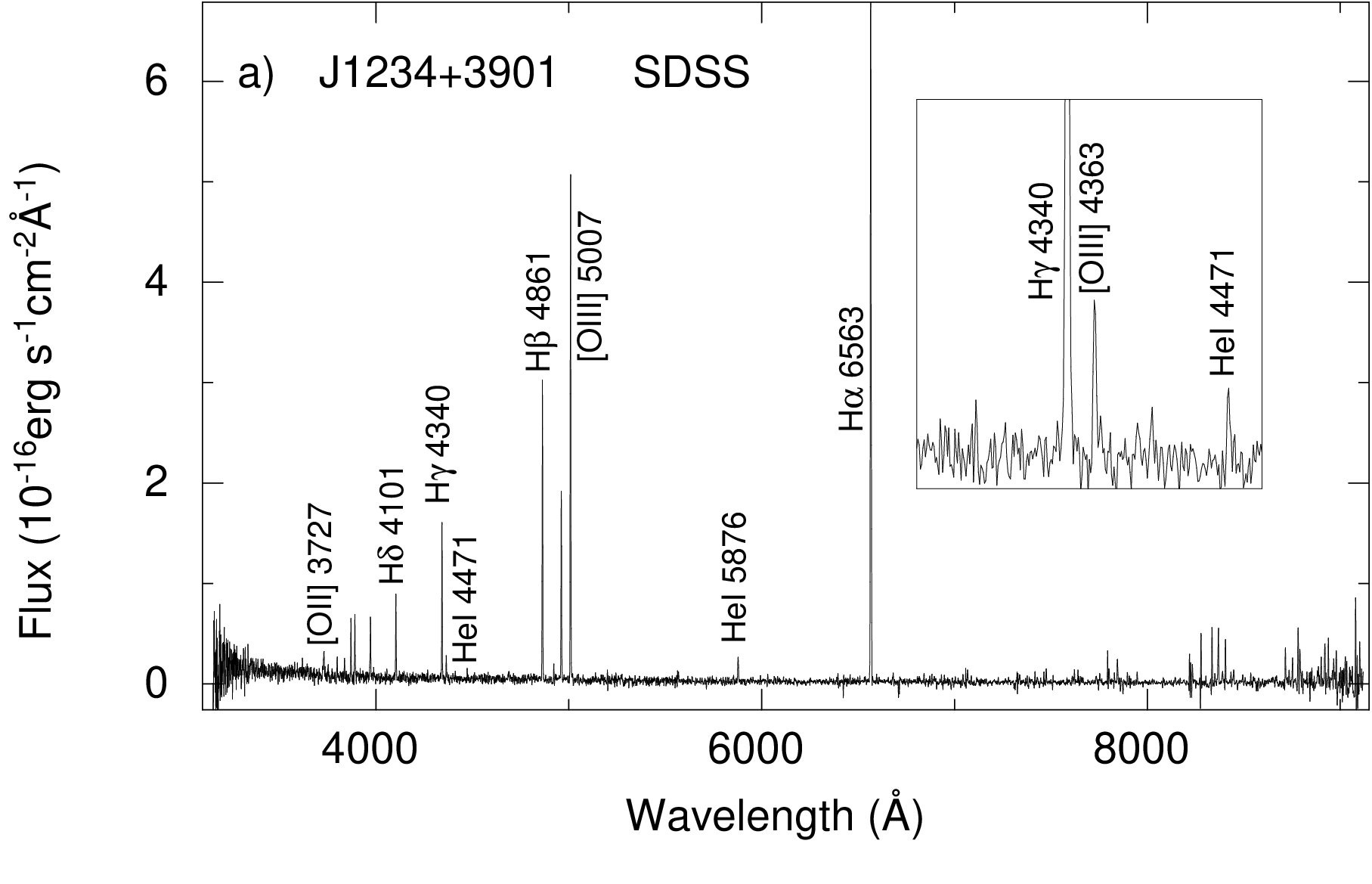}
\hspace{0.4cm}\includegraphics[angle=-90,width=0.48\linewidth]{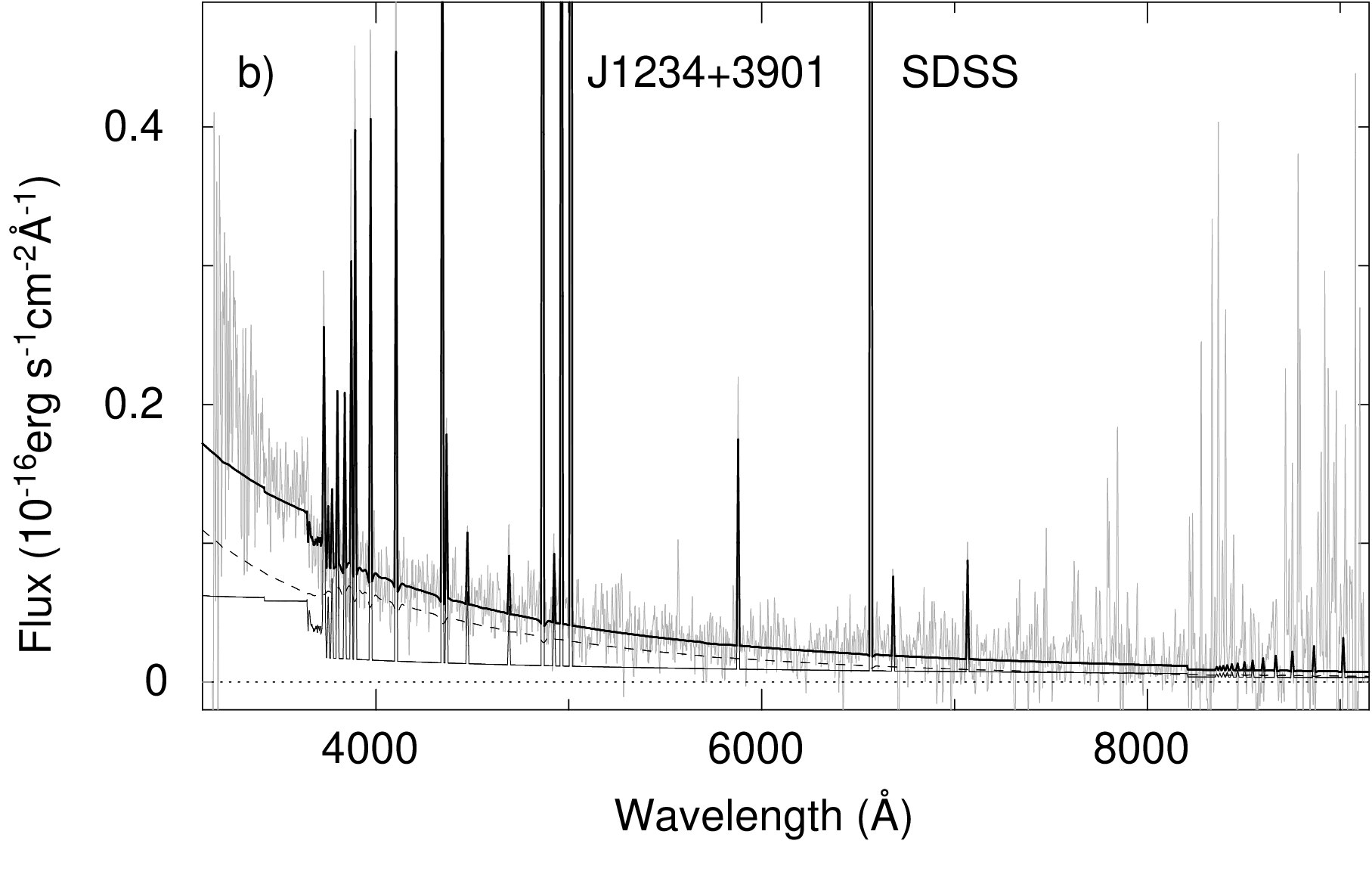}
}
\hbox{
\includegraphics[angle=-90,width=0.48\linewidth]{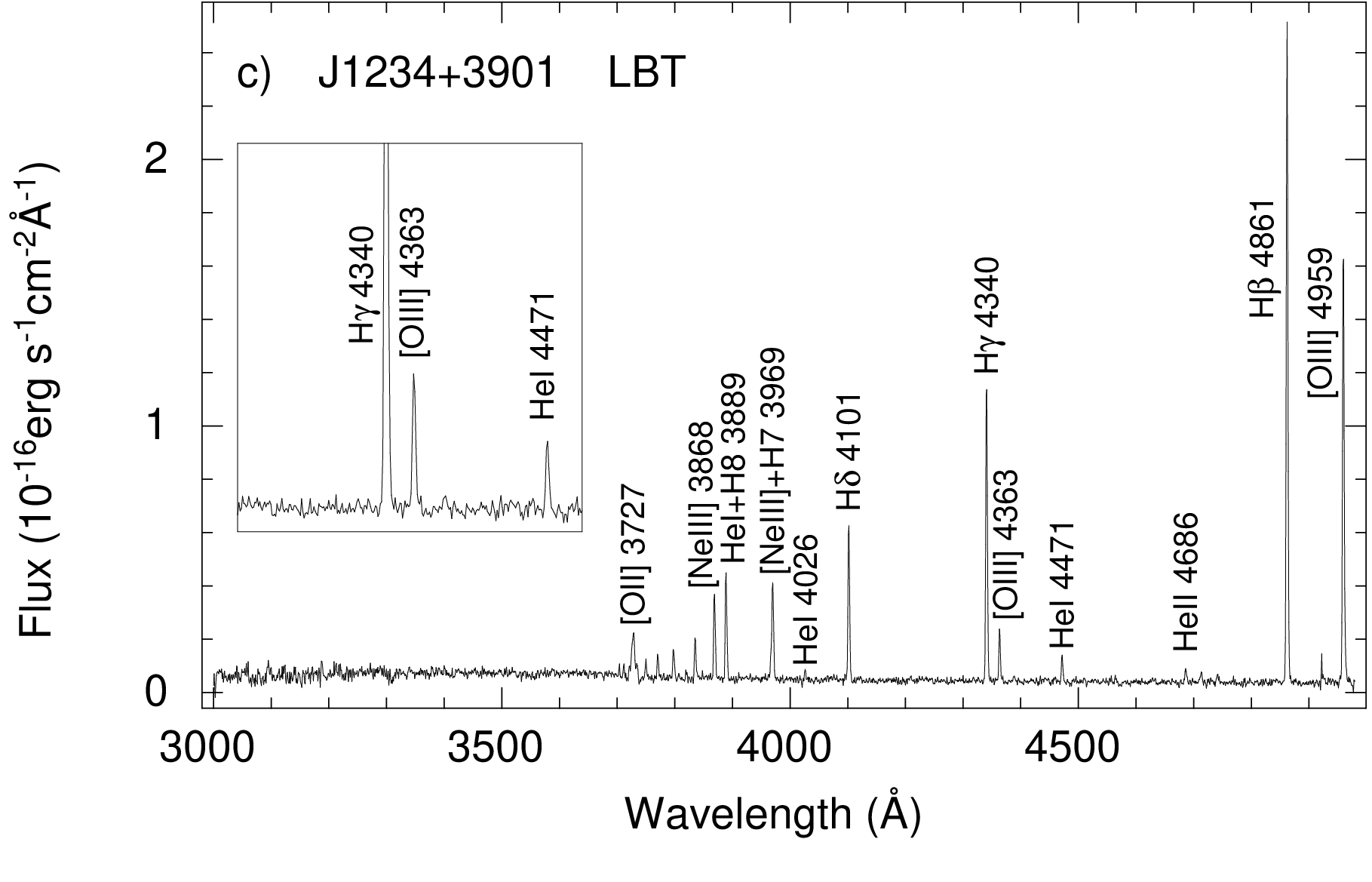}
\hspace{0.4cm}\includegraphics[angle=-90,width=0.48\linewidth]{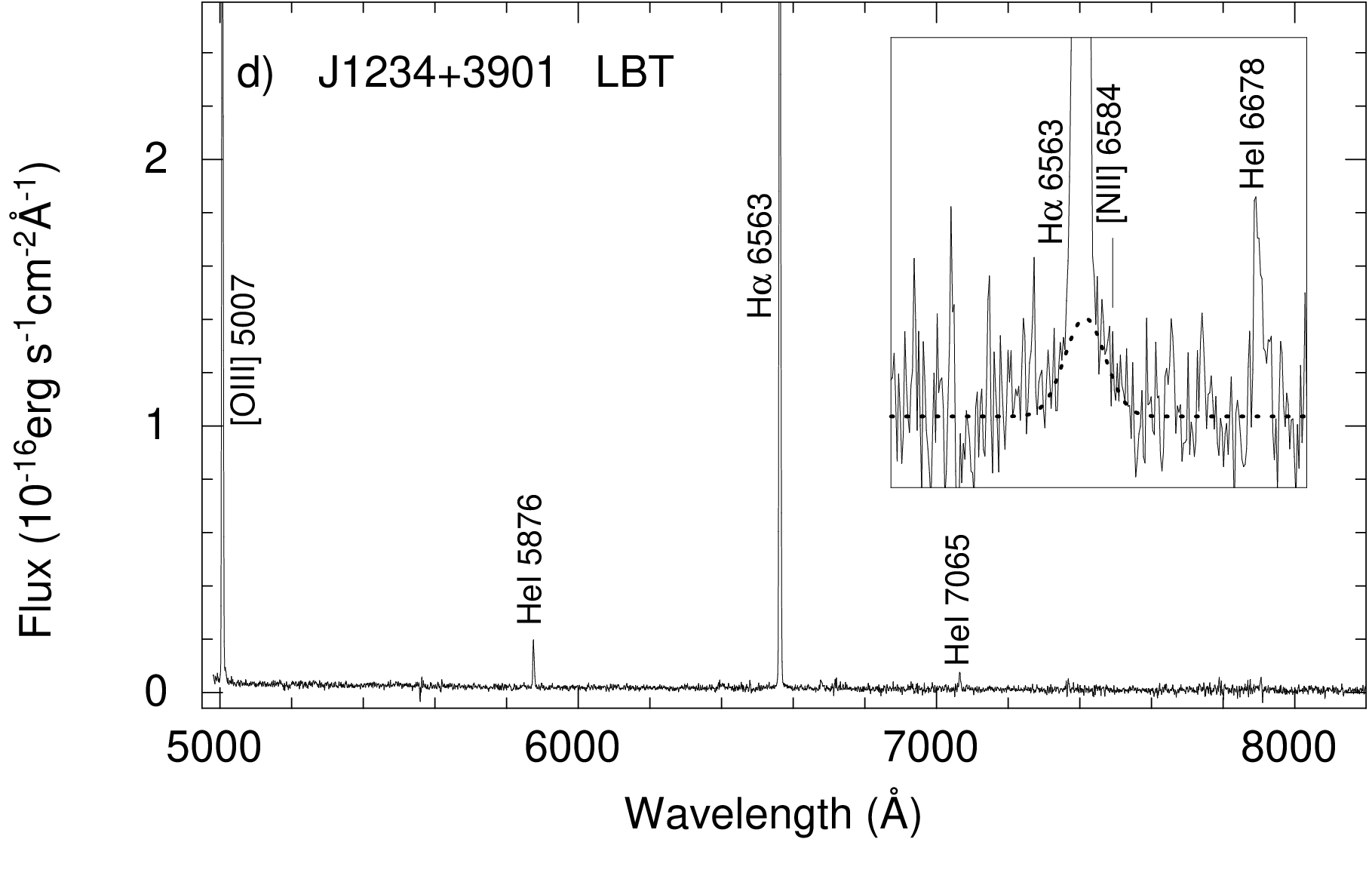}
}
\caption{{\bf a)} The rest-frame SDSS spectrum of J1234$+$3901.
{\bf b)} The rest-frame SDSS spectrum of J1234$+$3901 (grey line) on which is superposed the modelled spectral energy distribution
(SED). The thin solid and dashed lines represent modelled nebular and stellar SEDs
while the solid thick line shows their sum (total SED).
{\bf c)} and {\bf d)} The rest-frame LBT spectrum of J1234$+$3901.
Insets in {\bf a)}, {\bf c)} and {\bf d)} 
show expanded parts of spectral regions around the H$\gamma$ and 
H$\alpha$ emission lines for a better view of weak features.
Some emission lines in {\bf a)}, {\bf c)} and {\bf d)} are labelled. 
}
\label{fig1}
\end{figure*}

In this paper, we present LBT\footnote{The LBT 
is an international collaboration among institutions in the United States, 
Italy and Germany. LBT Corporation partners are: The University of Arizona on 
behalf of the Arizona university system; Istituto Nazionale di Astrofisica, 
Italy; LBT Beteiligungsgesellschaft, Germany, representing the Max-Planck 
Society, 
the Astrophysical Institute Potsdam, and Heidelberg University; The Ohio State 
University, and The Research Corporation, on behalf of The University of Notre 
Dame, University of Minnesota and University of Virginia.} spectroscopic
observations of a new SFG, J1234$+$3901 (the IAU designation is
SDSS J123415.69$+$390116.4), which can also be classified as a BCD. 
The galaxy was selected 
from the SDSS Data Release 14 (DR14) data base \citep{A18}, based on various 
emission line ratios, as one of the most promising candidates for being a very 
low metallicity SFG with high-excitation 
H~{\sc ii} regions (see the SDSS spectrum in Fig.~\ref{fig1}a). 
Its coordinates, redshift and other characteristics obtained from the 
photometric and spectroscopic data are presented in Table \ref{tab1}. For 
comparison, we also show in the Table similar data for the most 
metal-poor BCD known, J0811$+$4730 from \citet{I18a}.

\begin{table}
\caption{Comparison of observed characteristics of J0811$+$4730 and J1234$+$3901 \label{tab1}}
\begin{tabular}{lrr} \hline
Parameter                 & J0811$+$4730  &  J1234$+$3901       \\ \hline
R.A.(J2000)               &  08:11:52.12  &  12:34:15.70 \\
Dec.(J2000)               & +47:30:26.24  & +39:01:16.41 \\
  $z$                     &  0.04444         &  0.13297     \\
{\sl GALEX}  FUV, mag     &        ...       &   21.24$\pm$0.38      \\
{\sl GALEX}  NUV, mag     &        ...       &   22.17$\pm$0.57      \\
SDSS $g$, mag             &   21.37$\pm$0.05 &   21.92$\pm$0.06      \\
{\sl WISE}  $W1$, mag     &        ...       &   16.99$\pm$0.10      \\
{\sl WISE}  $W2$, mag     &        ...       &   16.16$\pm$0.17      \\
   $D_L$, Mpc$^{*}$              &        198       &       650     \\
 $M_g$, mag$^\dag$        & $-$15.41$\pm$0.06& $-$17.35$\pm$0.06     \\
log $L_g$/L$_{g,\odot}$$^\ddag$&     8.35$\pm$0.03&     9.12$\pm$0.03     \\
log $M_\star$/M$_\odot$$^{\dag\dag}$&   6.24$\pm$0.33&   7.13$\pm$0.30       \\
$M_\star$/$L_g$, M$_\odot$/L$_{g,\odot}$&  0.0078 & 0.0102       \\
$L$(H$\beta$), erg s$^{-1}$$^{**}$&(2.1$\pm$0.1)$\times$10$^{40}$&(4.9$\pm$0.3)$\times$10$^{40}$\\
SFR, M$_\odot$yr$^{-1}$$^{\ddag\ddag}$  &     0.48$\pm$0.02 &     1.08$\pm$0.08 \\
  12+logO/H$^{\dag\dag\dag}$              &6.979$\pm$0.019&7.035$\pm$0.026 \\
\hline
  \end{tabular}


\noindent$^{*}$Luminosity distance.

\noindent$^\dag$Corrected for Milky Way extinction.

\noindent$^\ddag$log of the SDSS $g$-band luminosity corrected for Milky Way extinction.

\noindent$^{\dag\dag}$Derived from the extinction- and aperture-corrected SDSS
spectrum.


\noindent$^{**}$Corrected for extinction and the SDSS spectroscopic aperture.

\noindent$^{\ddag\ddag}$Derived from the \citet{K98} relation using the extinction- 
and aperture-corrected H$\beta$ luminosity.

\noindent$^{\dag\dag\dag}$Oxygen abundances of J0811$+$4730 \citep{I18a} and
J1234$+$3901 (this paper) derived from the LBT spectra.

  \end{table}

The LBT observations of J1234$+$3901 and data reduction are described in 
Sect.~\ref{sec:observations}. We derive its element abundances in 
Sect.~\ref{sec:abundances}. Integrated characteristics of J1234$+$3901 are
presented in Sect.~\ref{sec:integr}. In Sect.~\ref{sec:highion} we discuss the 
origin of the hard ionizing radiation responsible for the strong He~{\sc ii} $\lambda$4686 
emission line. The 
possibility of Lyman continuum leakage in J1234$+$3901 is
considered in Sect.~\ref{sec:diagrams}.
We summarize our main results in Sect.~\ref{sec:conclusions}.

\section{LBT Observations and data reduction}\label{sec:observations}

We have obtained LBT long-slit spectrophotometric observations of J1234$+$3901 
on 6 June, 2018 in the twin binocular mode using the MODS1
spectrograph\footnote{This paper used data obtained with the MODS 
spectrographs built with
funding from NSF grant AST-9987045 and the NSF Telescope System
Instrumentation Program (TSIP), with additional funds from the Ohio
Board of Regents and the Ohio State University Office of Research.}.
As for MODS2, due to a technical problem with the CCD of the red channel,
observations were executed only with the blue channel. Spectra were 
obtained in the wavelength range 3200 -- 10000\AA\ with a 1.2 arcsec wide slit, 
resulting in a resolving power $R$ $\sim$ 2000.
The seeing during the observations was 0.6 arcsec.
The angular size of J1234$+$3901 as measured by the Full Width at Half 
Maximum of the Point Spread function is 1.28 arcsec. Thus the LBT slit width 
should include most of the light of the galaxy. We will discusss this issue 
more quantitatively in Section \ref{sec:integr}.

Four 900~s subexposures were obtained in the blue range separately with MODS1 
and MODS2, resulting in a total exposure time of 2$\times$3600~s, counting both 
spectrographs. In the red range, three 900~s subexposures were obtained with 
the MODS1, resulting in the total exposure of 2700~s.
The airmass during observations was in the range 1.05 -- 1.15.
Thus, the effect of atmospheric refraction is small 
for all subexposures \citep[see ][]{F82}. 

The spectrum of the spectrophotometric standard star 
GD~153 was obtained during the same night with a 5 arcsec wide slit 
for flux calibration and correction for telluric absorption in the
red part.
Additionally, calibration frames of biases, flats and comparison lamps 
were obtained during the same period with the same setups of MODS1 and MODS2.

The MODS Basic CCD Reduction package {\sc modsccdred}\footnote{http://www.astronomy.ohio-state.edu/MODS/Manuals/ MODSCCDRed.pdf} and 
{\sc iraf}\footnote{{\sc iraf} is distributed by the 
National Optical Astronomy Observatories, which are operated by the Association
of Universities for Research in Astronomy, Inc., under cooperative agreement 
with the National Science Foundation.} were used for bias subtraction, 
flat field correction, wavelength and flux calibration.
After these reduction steps, MODS1 and MODS2 subexposures in the blue part 
of the spectrum and MODS1 subexposures in the red part were 
combined. The one-dimensional spectrum 
of J1234$+$3901 extracted in a 1.2 arcsec aperture along the spatial axis 
is shown in Fig. \ref{fig1}c and \ref{fig1}d. Strong emission lines are 
present in the spectrum, suggesting active star formation.
In particular, a strong [O~{\sc iii}] $\lambda$4363 emission line is 
detected with a signal-to-noise ratio much higher than in the SDSS spectrum
(compare insets in Fig.~\ref{fig1}a and Fig.~\ref{fig1}c), allowing  
reliable abundance determinations. The another notable feature detected
in both the SDSS and LBT spectra is the He~{\sc ii} $\lambda$4686 emission line
(Fig.~\ref{fig1}c), implying the presence of hard ionizing radiation with energy
above 4 Ryd. We also note that the weak [N~{\sc ii}]$\lambda$6584 emission line
is not detected in the LBT spectrum because of insufficient signal-to-noise
ratio (see inset in Fig.~\ref{fig1}d). This fact precludes the determination
of the nitrogen abundance. On the other hand, weak broad wings of H$\alpha$
are present (see inset in Fig.~\ref{fig1}d), indicating fast moving 
ionized gas in J1234$+$3901 with a velocity dispersion 
$\sigma$ $\sim$ 850 km s$^{-1}$. However, this estimate is somewhat uncertain 
because of the low intensity of the broad H$\alpha$ emission line and a noisy
continuum.

\begin{table}
\caption{Extinction-corrected emission-line flux ratios$^{*}$ \label{tab2}}
\begin{tabular}{lrr} \hline
Line& \multicolumn{1}{c}{SDSS}  & \multicolumn{1}{c}{LBT}   \\ \hline
3187.74 He {\sc i}              &\multicolumn{1}{c}{...}    &  1.93$\pm$0.24\\
3703.30 H16                     &\multicolumn{1}{c}{...}    &  1.28$\pm$0.17\\
3711.97 H15                     &\multicolumn{1}{c}{...}    &  2.63$\pm$0.40\\
3721.94 H14                     &\multicolumn{1}{c}{...}    &  0.95$\pm$0.11\\
3727.00 [O {\sc ii}]            & 11.74$\pm$2.38& 12.76$\pm$0.45\\
3734.37 H13                     &\multicolumn{1}{c}{...}    &  1.51$\pm$0.13\\
3750.15 H12                     &\multicolumn{1}{c}{...}    &  4.62$\pm$0.42\\
3770.63 H11                     &\multicolumn{1}{c}{...}    &  5.72$\pm$0.43\\
3797.90 H10                     &  6.39$\pm$2.45&  6.75$\pm$0.43\\
3819.64 He {\sc i}              &\multicolumn{1}{c}{...}    &  1.08$\pm$0.11\\
3835.39 H9                      &  7.01$\pm$2.33&  7.87$\pm$0.41\\
3868.76 [Ne {\sc iii}]          & 15.79$\pm$2.56& 12.77$\pm$0.44\\
3889.00 He {\sc i}+H8           & 20.76$\pm$3.20& 17.70$\pm$0.63\\
3968.00 [Ne {\sc iii}]+H7       & 21.59$\pm$3.27& 20.16$\pm$0.71\\
4026.19 He {\sc i}              &\multicolumn{1}{c}{...}    &  1.59$\pm$0.14\\
4101.74 H$\delta$               & 25.79$\pm$3.38& 26.03$\pm$0.85\\
4227.20 [Fe {\sc v}]            &\multicolumn{1}{c}{...}    &  0.66$\pm$0.21\\
4340.47 H$\gamma$               & 45.45$\pm$4.60& 46.37$\pm$1.40\\
4363.21 [O {\sc iii}]           &  6.94$\pm$1.88&  7.73$\pm$0.29\\
4471.48 He {\sc i}              &  3.19$\pm$1.50&  3.82$\pm$0.19\\
4685.94 He {\sc ii}             &  3.03$\pm$1.51&  2.42$\pm$0.20\\
4712.00 [Ar {\sc iv}]+He {\sc i}&\multicolumn{1}{c}{...}    &  1.87$\pm$0.17\\
4740.20 [Ar {\sc iv}]           &\multicolumn{1}{c}{...}    &  1.73$\pm$0.17\\
4861.33 H$\beta$                &100.00$\pm$7.58&100.00$\pm$2.89\\
4958.92 [O {\sc iii}]           & 64.95$\pm$5.48& 68.40$\pm$2.01\\
5006.80 [O {\sc iii}]           &187.06$\pm$12.4&197.00$\pm$5.74\\
5875.60 He {\sc i}              & 10.66$\pm$1.91& 10.09$\pm$0.50\\
6562.80 H$\alpha$               &272.07$\pm$18.9&271.29$\pm$8.48\\
6678.10 He {\sc i}              &  4.09$\pm$1.23&  2.43$\pm$0.33\\
7065.30 He {\sc i}              &  5.10$\pm$1.30&  3.85$\pm$0.41\\ \\
$C$(H$\beta$)$^{\dag}$         &\multicolumn{1}{c}{0.000$\pm$0.083}&\multicolumn{1}{c}{0.000$\pm$0.037}\\
$F$(H$\beta$)$^{\ddag}$         &\multicolumn{1}{c}{10.93$\pm$0.58}&\multicolumn{1}{c}{9.74$\pm$0.28}\\
EW(H$\beta$)$^{**}$          &\multicolumn{1}{c}{276.0$\pm$14.5}&\multicolumn{1}{c}{242.2$\pm$1.0}\\
EW(abs)$^{**}$               &\multicolumn{1}{c}{0.4$\pm$2.4}&\multicolumn{1}{c}{2.7$\pm$0.4}\\
\hline
  \end{tabular}

\hbox{$^{*}$in units 100$\times$$I(\lambda)$/$I$(H$\beta$).} 

\hbox{$^{\dag}$Extinction coefficient, derived from the observed hydrogen} 

\hbox{\,~Balmer decrement.}

\hbox{$^{\ddag}$Observed flux in units of 10$^{-16}$ erg s$^{-1}$ cm$^{-2}$.}

\hbox{$^{**}$Equivalent width in \AA.}

  \end{table}

\begin{table}
\caption{Electron temperatures, electron number densities 
and heavy element abundances \label{tab3}}
\begin{tabular}{lccccc} \hline
Property                             &SDSS&LBT          \\ \hline
$T_{\rm e}$(O {\sc iii}), K          &21100$\pm$3700&   21900$\pm$600       \\
$T_{\rm e}$(O {\sc ii}), K           &15600$\pm$2500&   15600$\pm$400       \\
$N_{\rm e}$(S {\sc ii}), cm$^{-3}$    &10$\pm$10     &10$\pm$10         \\ \\
O$^+$/H$^+$$\times$10$^6$            &0.939$\pm$0.425 &1.032$\pm$0.077 \\
O$^{2+}$/H$^+$$\times$10$^5$          &0.969$\pm$0.382 &0.950$\pm$0.064 \\
O$^{3+}$/H$^+$$\times$10$^6$          &0.236$\pm$0.162 &0.313$\pm$0.054 \\
O/H$\times$10$^5$                   &1.086$\pm$0.385 &1.085$\pm$0.065 \\
12+log(O/H)                         &7.036$\pm$0.154 &7.035$\pm$0.026     \\ \\
Ne$^{2+}$/H$^+$$\times$10$^6$        &1.770$\pm$0.691 &1.322$\pm$0.087 \\
ICF(Ne)                             &1.046 &1.053 \\
Ne/H$\times$10$^6$                  &1.852$\pm$0.775 &1.392$\pm$0.100 \\
log(Ne/O)                           &$-$0.768$\pm$0.238~~~&$-$0.892$\pm$0.041~~~\\ \\
\hline
  \end{tabular}
  \end{table}

The observed emission-line fluxes and their errors in both the SDSS and LBT 
spectra were measured using the  {\sc iraf splot} routine. 
Following \citet{ITL94} they were corrected for extinction and underlying 
stellar absorption, derived from the observed decrement of the hydrogen Balmer 
emission lines. The equivalent widths of the underlying stellar Balmer 
absorption lines are assumed to be the same for each line. 
The fluxes corrected for extinction (all lines) and underlying stellar 
absorption (hydrogen lines) are shown 
in Table \ref{tab2} for both the SDSS and LBT observations. 
The Table also includes the extinction 
coefficient $C$(H$\beta$), the observed H$\beta$ emission-line flux 
$F$(H$\beta$), the rest-frame equivalent width EW(H$\beta$) of the H$\beta$ 
emission line, and the equivalent width of the Balmer absorption lines.
Within the errors, the derived extinction $C$(H$\beta$) $\sim$ $A(V)$/2.1 \citep{A84}  
is zero (Table \ref{tab2}). This null value is consistent with the low Milky Way 
extinction $A(V)$ = 0.042 mag (NASA Extragalactic Database) and implies a very low
internal galaxy extinction, likely due to the very low metallicity
of J1234$+$3901. A similar conclusion  can be made for J0811$+$4730 with
the lowest luminosity-weighted oxygen abundance known \citep{I18a}. Its internal extinction 
$A(V)$ = 0.167 mag is smaller than the Milky Way extinction $A(V)$ = 0.180 mag.
Low $A(V)$ $<$ 0.5 mag are typical for low-redshift compact SFGs 
\citep{I14,G17}.
We finally note that the EW(H$\beta$) in J1234$+$3901 is high, $\sim$~276\AA\ in the 
SDSS spectrum and $\sim$~242\AA\ in the LBT spectrum, implying that 
its optical emission is dominated by radiation from a very young starburst, with
age $\sim$ 3 Myr.

\section{Heavy element abundances}\label{sec:abundances}

The procedures described by \citet{I06a} are used to determine heavy
element abundances from the SDSS and LBT spectra. 
The temperature $T_{\rm e}$(O~{\sc iii}) is calculated from the 
[O~{\sc iii}] $\lambda$4363/($\lambda$4959 + $\lambda$5007) emission-line flux 
ratio. It is used to derive the abundances of O$^{3+}$, O$^{2+}$ and Ne$^{2+}$.
The abundance  of O$^{+}$ is derived with the electron
temperature $T_{\rm e}$(O~{\sc ii}), using the relations of \citet{I06a}
between $T_{\rm e}$(O~{\sc ii}) and $T_{\rm e}$(O~{\sc iii}). These relations 
have been obtained from the ionization-bounded H~{\sc ii} region models of 
\citet{SI03}, assuming the low-density limit, when collisional de-excitation
from upper levels of forbidden transitions is unimportant \citep[e.g. ][]{A84}, 
and adopting the stellar
evolution models of \citet{M94} and the stellar  atmosphere models of 
\citet*{S02}. To check the integrity of our 
$T_{\rm e}$(O~{\sc ii}) -- $T_{\rm e}$(O~{\sc iii}) relations, we have also 
derived relations using as input the {\sc cloudy} v17.01 
H~{\sc ii} region models of \citet{F17}. The resulting relations are very 
similar to the ones obtained by  \citet{I06a}. 
We note that in the particular case of J1234$+$3901,
the fraction of O$^+$ ions is one order of magnitude lower than that of
O$^{2+}$ ions because of the very high O$_{32}$ ratio 
(see Section \ref{sec:diagrams}).
Therefore, uncertainties of $T_{\rm e}$(O~{\sc ii}) of $\sim$ 10 per cent will
result in uncertainties of oxygen abundances $\la$ $\pm$ 0.01 dex.
The [S~{\sc ii}] $\lambda$6717, $\lambda$6731 emission lines are not detected 
and the [O~{\sc ii}]$\lambda$3726, $\lambda$3729 emission lines are not resolved
in both the SDSS and LBT spectra. Therefore, the electron number density 
$N_{\rm e}$(S~{\sc ii}) cannot be determined from the spectra. We adopted it 
to be 10 cm$^{-3}$. The precise adopted value does not influence the determinations
of oxygen and neon abundances because they do not depend sensitively on 
$N_{\rm e}$ for typical electron number densities of $\la$ 10$^3$ cm$^{-3}$ in 
H~{\sc ii} regions \citep[e.g. ][]{A84}.

The total oxygen abundance is derived as follows:
\begin{equation}
\frac{\rm O}{\rm H} = \frac{{\rm O}^++{\rm O}^{2+}+{\rm O}^{3+}}{{\rm H}^+}, 
\label{OH} 
\end{equation}
where the abundances of ions O$^+$, O$^{2+}$, O$^{3+}$ are obtained using 
the relations of \citet{I06a}. For neon, we also use
the relations of \citet{I06a} to derive the Ne$^{2+}$ abundance, the 
ionization correction factor ICF(Ne) and the total Ne abundance.

The electron temperatures, electron number densities, ionic abundances,
ionization correction factors and total O and Ne abundances are presented 
in Table~\ref{tab3}. The electron temperatures $T_{\rm e}$(O~{\sc iii}) of 
21900 $\pm$ 600 K derived from the LBT spectrum and of 21100 $\pm$ 3700 K 
derived from the SDSS spectrum are high. This is a consequence of the very low 
metallicity of J1234$+$3901.

The nebular oxygen abundance of
12+logO/H = 7.035$\pm$0.026 derived from the LBT spectrum is among the lowest 
found for SFGs. A similar value was obtained from the SDSS spectrum, although 
with a higher error because of the lower signal-to-noise ratio.
The Ne/O abundance ratio for this galaxy (Table \ref{tab3}), is similar to 
that in other low-metallicity SFGs \citep[e.g., ][]{I06a}.

\section{Integrated characteristics of J1234$+$3901}
\label{sec:integr}

The stellar mass, absolute magnitudes and luminosities of J1234$+$3901 are 
derived adopting the luminosity distance $D_L$ = 650 Mpc, obtained from
the galaxy redshift for the cosmological parameters 
$H_0$ = 67.1 km s$^{-1}$ Mpc$^{-1}$, $\Omega_m$ = 0.318, $\Omega_\Lambda$ = 
0.682 \citep{P14} and assuming a flat geometry.

The absolute SDSS $g$ magnitude, corrected for the Milky Way 
extinction is $M_g$ = $-$17.35 mag (Table \ref{tab1}). Thus, J1234$+$3901
is the most luminous BCD known with 12~+~logO/H $\la$ 7.3.  
Like most other extremely metal-deficient SFGs, it strongly deviates from the luminosity -- metallicity relation defined by the bulk of the SDSS SFGs,
\citep[e.g. see fig. 5 in ][]{I18a}.

The star-formation rate (SFR) in J1234$+$3901, derived from the SDSS H$\alpha$ 
luminosity (Table \ref{tab2})
using the \citet{K98} calibration 
is equal to 1.1~M$_\odot$~yr$^{-1}$.  

The stellar mass of J1234$+$3901 is determined from fitting the 
spectral energy distribution (SED). According to the SDSS database, 
the Petrosian radius $R_{90}$ within which 
90 per cent of the galaxy's light in the SDSS $r$ band is contained, is equal 
to 1.06 arcsec. Therefore, for SED fitting, we have used the SDSS spectrum 
obtained with a round 2 arcsec aperture which contains most of the galaxy's 
light, instead of the LBT spectrum obtained with a 
narrow 1.2 arcsec slit. Some light is missed in the case of the latter,  
as evidenced by the lower observed H$\beta$ flux (Table \ref{tab2}).
The difference in the EW(H$\beta$)s of the LBT and SDSS spectra can also be
explained by an aperture effect. Assuming that the central ionizing cluster is 
inside both the SDSS and LBT apertures, but some H$\beta$ flux is outside the 
LBT slit, we should correct the EW(H$\beta$) of 242.2 \AA\ in the LBT spectrum 
by a factor 10.93/9.74, the ratio of the H$\beta$ fluxes in the SDSS and LBT 
spectra. We obtain 271.8 \AA\ which agrees well, within the errors, with the
observed value of the EW(H$\beta$) in the SDSS spectrum.

\begin{figure}
\centering{
\includegraphics[angle=-90,width=0.99\linewidth]{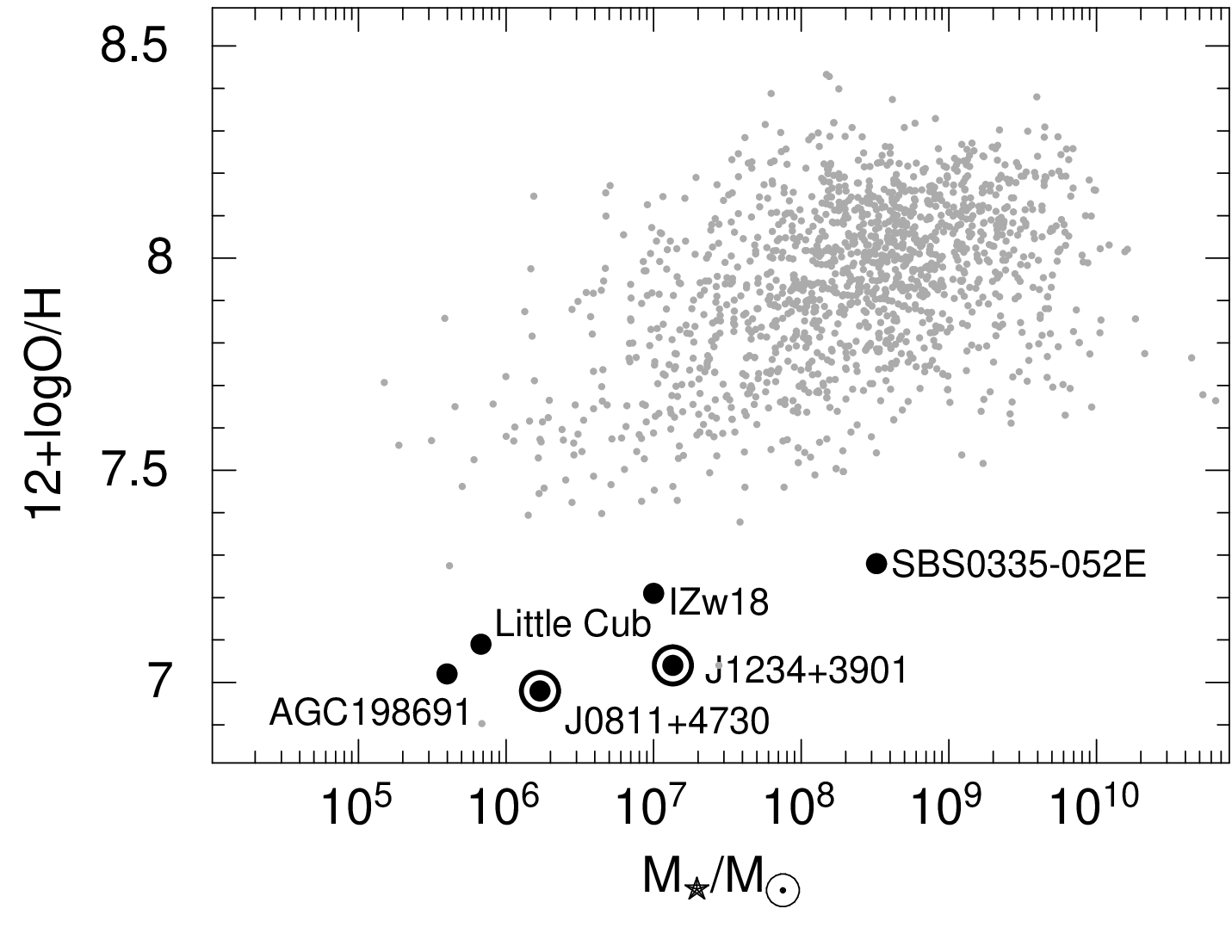}
}
\caption{The stellar mass - metallicity relation. J0811$+$4730 and J1234$+$3901 
are shown by labelled encircled filled circles and some other extremely
low-metallicity compact SFGs from \citet[][I~Zw~18, SBS~0335$-$052E]{H15}, 
\citet[][A198691]{H16} and \citet[][Little Cub]{H17} are represented by 
labelled filled circles. A subsample of 1397 SFGs from the SDSS
DR14 is represented by grey dots. 
This subsample includes only 
compact galaxies with an [O~{\sc iii}]$\lambda$4363 emission line in their 
spectra, measured with an accuracy better than 25 per cent. It is a subsample of  
the entire sample of compact SFGs selected from the SDSS data 
base which includes $\sim$ 20,000 objects. The selection criteria for the entire sample 
were a compact structure with an angular diameter $<$ 6 arcsec and a high 
equivalent width EW(H$\beta$) $\ga$ 10\AA\ of the H$\beta$ emission line, 
indicating active star formation.}
\label{fig2}
\end{figure}

To carry out the SED fitting, we calculated
a grid of SEDs for instantaneous burst models in the age and heavy element
mass fraction ranges of 1 Myr -- 10 Gyr and $Z$ = 0.0004 -- 0.02, respectively.
We use {\sc starburst99} models \citep{L99} 
and adopt stellar evolution models by \citet{G00},
stellar atmosphere models by \citet*{L97} and the \citet{S55} initial mass function (IMF) with lower and upper mass limits of 0.1 M$_\odot$ and 100 M$_\odot$,
respectively.

Because the equivalent width of the H$\beta$ emission line is high 
(Table \ref{tab2}), the contribution of the nebular continuum should be 
taken into account in the SED fitting, in addition to the stellar emission.
The nebular continuum was calculated adopting the electron temperature
derived from the galaxy emission-line spectrum. The fraction of the nebular
continuum flux in the total continuum flux near the H$\beta$ emission line 
is determined by the ratio of the observed equivalent width EW(H$\beta$) to
the H$\beta$ equivalent width of 900 -- 1100\AA\ (the precise value depends on the electron
temperature) for pure nebular emission.
The star-formation history in J1234$+$3901 was approximated 
by a recent short burst at age $t_{\rm b}$ $<$ 10 Myr and a prior 
continuous star formation with a constant SFR during the time interval
$t_2$ -- $t_1$ with $t_1$, $t_2$ $>$ 10 Myr and $t_2$ $>$ $t_1$. Finally, the total 
(stellar+nebular) SED is scaled to the observed continuum flux near H$\beta$.
We use a Monte Carlo method with $\chi^2$ minimization, varying $t_{\rm b}$,
$t_1$, $t_2$ and the mass fraction of stellar populations formed during the burst
and after the age of 10 Myr, and aiming to obtain the best agreement between the
modelled and observed continuum in the entire wavelength range of the SDSS 
spectrum. Additionally, the observed EW(H$\beta$) and EW(H$\alpha$) have to 
simultaneously be reproduced within 10 per cent of their values by the best model.
More details on the SED fitting procedure can be found e.g. in \citet{I18a}.

The SDSS spectrum superposed by the modelled stellar, nebular and total
SEDs is shown in Fig.~\ref{fig1}b. Despite the noisy data, the model SED 
reproduces reasonably well the observed SDSS spectrum, including the region 
shortward of the Balmer jump at $\lambda$ $\la$ 3660\AA. We obtain a low stellar
mass for J1234$+$3901, $M_\star$ = 10$^{7.13\pm0.30}$~M$_\odot$ (Table~\ref{tab1}). 
This yields in turn a very high specific star formation rate sSFR of 
$\sim$ 100 Gyr$^{-1}$, indicative of very active ongoing star formation.

In Fig. \ref{fig2} we show the mass -- metallicity relation for 
low-redshift SFGs. It is seen that both J0811$+$4730 and J1234$+$3901 strongly
deviate from the general relation defined by the bulk of the compact SFGs in 
the SDSS DR14 (grey dots) with oxygen abundances determined reliably by the 
direct $T_{\rm e}$ method. These two galaxies 
are 3 -- 5 times more metal-poor for their stellar masses.
They are also more extreme compared to other 
very metal-deficient SFGs found in the literature and shown by filled circles.

Using $M_\star$ and deriving the $g$-band luminosity from the relation
\begin{equation}
\log \frac{L_g}{{\rm L}_{g,\odot}}=0.4\log (M_{g,\odot} - M_g), \label{eq:Lg}
\end{equation}
where $M_{g,\odot}$ = 5.45 is the absolute $g$-band magnitude of the Sun 
\citep{B03}, we obtain the very low mass-to-luminosity ratio of $\sim$ 0.01 
(in solar units). This is more than one order of magnitude lower than 
the average value for SDSS SFGs, but similar to the one derived for the 
lowest-metallicity BCD known, J0811$+$4730 \citep[Table~\ref{tab1}, ][]{I18a}. 
This $M_\star$/$L_g$ value is consistent with the value
for a stellar population with age $<$ 10 Myr 
\citep{L99,L14,I18a}, indicating that the emission
of J1234$+$3901 in the optical range is dominated by young massive stars
with a negligible contribution of older stars, and possibly supporting the idea
that this SFG is a young galaxy having formed most of its stars only very 
recently.

\section{The origin of the hard ionizing radiation in J1234$+$3901}
\label{sec:highion}

An important feature of the J1234$+$3901 spectrum is the presence of a strong
narrow nebular He~{\sc ii} $\lambda$4686 emission line (Table~\ref{tab2}). 
A similarly strong He~{\sc ii} $\lambda$4686 emission line was detected by 
\citet{I18a} in the spectrum of the lowest-metallicity BCD known J0811$+$4730. 
This line is seen quite often in low-metallicity SFGs 
\citep[e.g. ][]{TI05,SB12}, but very rarely with a flux above 2 per cent that of
H$\beta$, as is the case for the two above BCDs. Its origin remains
unclear. Several mechanisms have been proposed for the creation of hard 
ionizing radiation with energy above 4 Ryd, responsible for the production of 
He$^{2+}$ ions followed by recombination He~{\sc ii} emission.
In particular, \citet*{I12b} have examined X-ray emission from 
AGN and high-mass X-ray binaries.  Although these cannot be ruled out as sources of hard ionizing photons, they are considered to be unlikely for the majority of low-mass galaxies with active ongoing star formation.
The most favoured mechanisms are extreme UV (EUV) radiation from hot 
massive stars and from relatively fast radiative shocks, as discussed below.

\subsection{Stellar radiation} \label{stellar}

At the moment, the role of stellar radiation in the production of He~{\sc ii}
emission remains controversial. While e.g. \citet{SB12}, based on the 
{\sc starburst99} population synthesis models, have suggested that hard 
radiation is produced by a Wolf-Rayet stellar population in young starbursts 
with age $<$ 4 -- 5 Myr, \citet*{GIT00} and \citet{TI05} have argued for another
mechanism, since no WR features were detected in most of the low-metallicity 
He~{\sc ii}-emitting galaxies with 12~+~logO/H~$<$~8.0.

One of the most studied He~{\sc ii} emitters is the BCD 
SBS~0335$-$052E with an oxygen abundance 12 + logO/H = 7.30. This galaxy is 
also the lowest-metallicity SFG with detected [Ne~{\sc v}] emission which
requires the presence of even harder radiation, with energy above 7 Ryd.
\citet*{I01}, \citet{TI05} and \citet{I06b} have discussed the origin of
He~{\sc ii} and [Ne~{\sc v}] emission in SBS~0335$-$052E. It was found
that the He~{\sc ii} emission is spatially distinct from its hydrogen H$\alpha$ 
and H$\beta$ emission and broader, suggesting that the main source of hard 
radiation in that BCD is not stars but more likely radiative shocks.
Observations of [Ne~{\sc v}] emission in other SFGs 
\citep{I04,I12b} support and strengthen the radiative shock hypothesis.

Recently \citet{K18} have reconsidered the origin of the 
He~{\sc ii} $\lambda$4686 emission in SBS~0335$-$052E. 
To study the role of stellar radiation in producing He~{\sc ii} emission, 
they adopted a set of stellar population synthesis models by \citet{E17} called 
Binary Population and Spectral Synthesis or {\sc bpass} v2.1 models. These are 
different from e.g. the {\sc starburst99} models \citep{L99,L14} in that they 
include binary stellar evolution. Those authors found that the He~{\sc ii} 
emission of SBS~0335$-$052E can only be produced by either single, rotating 
metal-free stars or a binary population with $Z$ $\sim$ 10$^{-5}$ and a 
’top-heavy’ IMF. A difference between the assumed stellar and 
observed interstellar medium (ISM) 
metallicities of several orders of magnitude does not appear to be reasonable,
even for the most metal-deficient galaxies known, J0811$+$4730 and J1234$+$3901.
Furthermore, \citet{K18} did not discuss the presence of
[Ne~{\sc v}] emission
in SBS~0335$-$052E which requires an even harder ionizing radiation than
He~{\sc ii} emission.

Below we consider the lowest-metallicity {\sc bpass} v2.1 stellar models 
to assess the role these
models play in the production of He~{\sc ii} emission in the particular cases of 
J0811$+$4730 and J1234$+$3901. 

\begin{figure}
\centering{
\includegraphics[angle=-90,width=0.90\linewidth]{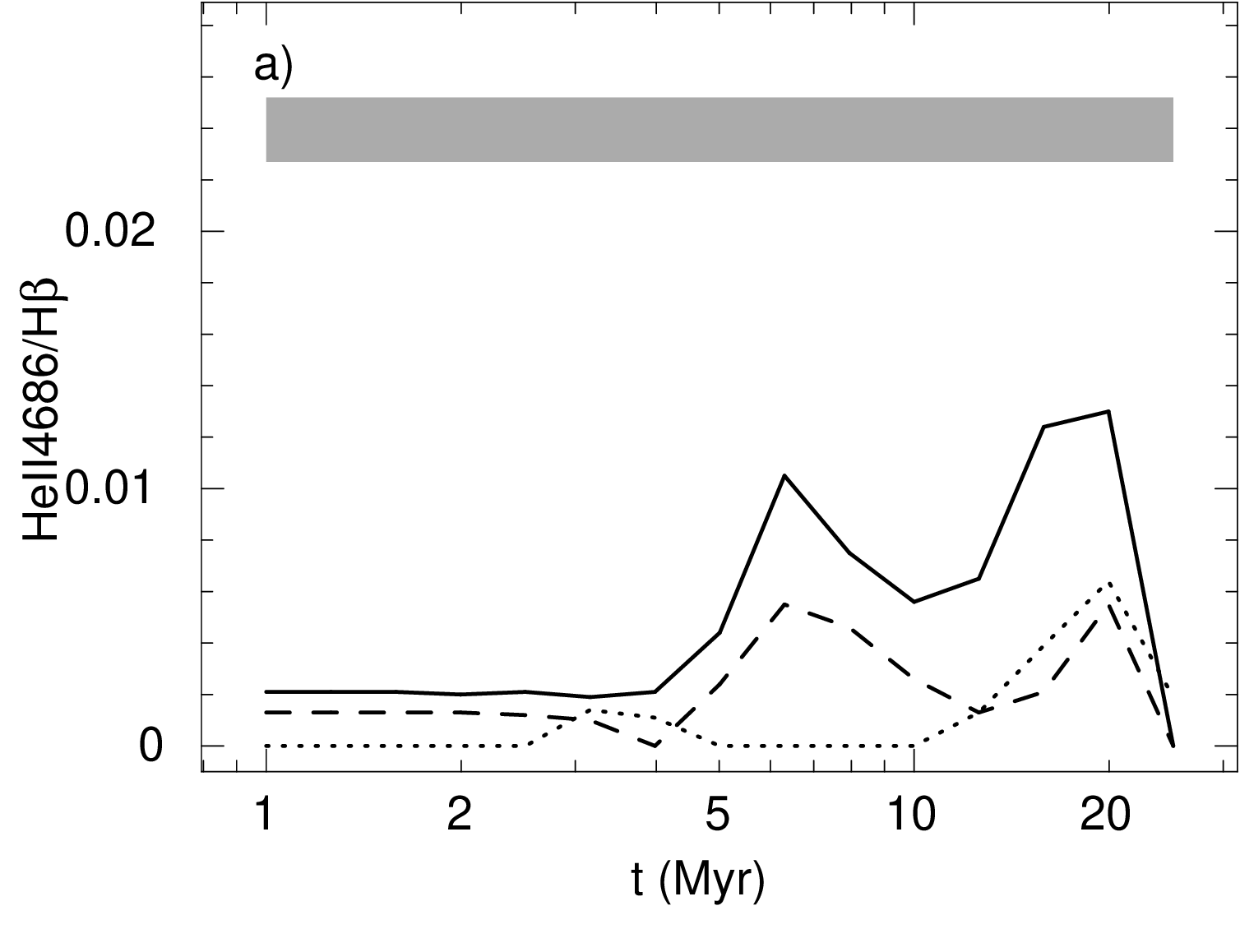}
\includegraphics[angle=-90,width=0.90\linewidth]{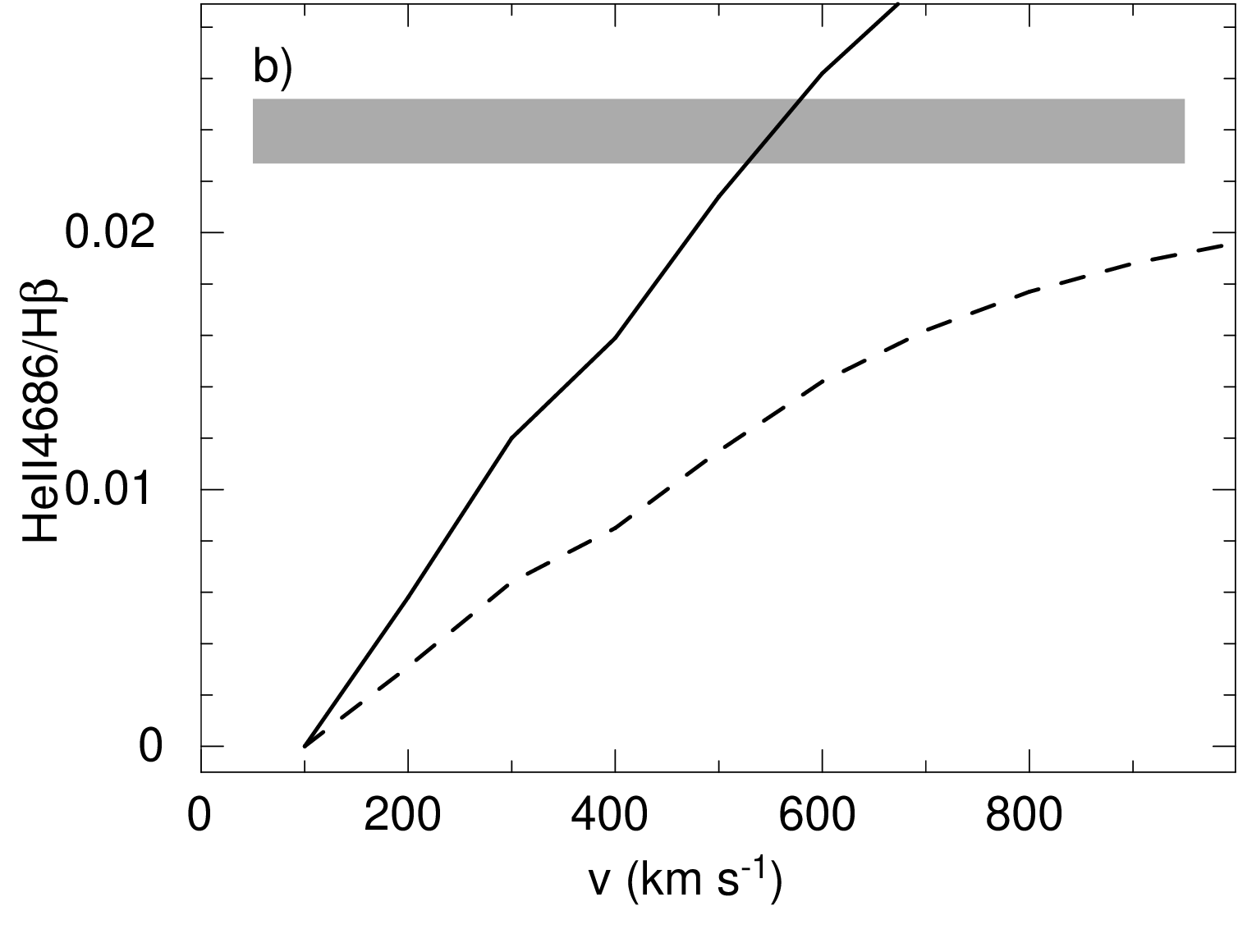}}
\caption{{\bf a)} Dependence of the He~{\sc ii} 4686/H$\beta$ emission-line
ratio on the age of the instantaneous burst in H~{\sc ii} region
models in which the source of ionizing photons is stellar. The solid, dashed and dotted 
lines show the predictions of the {\sc bpass} ionizing radiation models with stellar heavy element mass
fractions equal to 10$^{-5}$, 10$^{-4}$ and 10$^{-3}$, respectively. 
{\bf b)} Dependence of the He~{\sc ii} 4686/H$\beta$ emission-line
ratio on the shock velocity $v$ in composite models including ionizing radiation
from a {\sc starburst99} population synthesis model of an instantaneous burst
with age 2 Myr and a heavy element mass fraction $Z$ = 10$^{-3}$, and
radiative shock models. The solid and dashed lines show the model predictions for the cases when the production rate of ionizing
photons by the shock is respectively 10 per cent and 5 per cent 
that of stellar ionizing radiation. In both panels, the 
shaded region indicates the range of the observed He~{\sc ii} 4686/H$\beta$ 
flux ratios in J0811$+$4730 
and J1234$+$3901. For all models, a nebular oxygen abundance 12~+~logO/H~=~7.0 
(both panels) and an ionization parameter $U$~=~10$^{-2.3}$ (second panel)
have been adopted.
}
\label{fig3}
\end{figure}

We use the {\sc cloudy} v17.01 model calculations \citep{F17}
in conjunction with the {\sc bpass} v2.1 stellar models 
with $Z$ = 10$^{-3}$, 10$^{-4}$ and 10$^{-5}$, the ionization parameter $U$ in
the range 10$^{-3.0}$ -- 10$^{-1.6}$, and an ISM oxygen abundance 
12 + logO/H = 7.0, to calculate the temporal evolution of 
the He~{\sc ii} 4686/H$\beta$ emission line ratio (Fig.~\ref{fig3}a). 

It is seen
that binary evolution considerably increases the He~{\sc ii} $\lambda$4686
emission line intensity, producing two maxima at $t$ = 6 and 20 Myr.
However, only models with $Z$ = 10$^{-5}$
are able to produce He~{\sc ii} $\lambda$4686/H$\beta$ $>$ 1 per cent. 
But even these peak values are considerably 
lower than those observed in J0811$+$4730 and J1234$+$3901 (shaded horizontal 
region). We note that the He~{\sc ii} $\lambda$4686/H$\beta$ flux ratio is 
nearly independent of $U$.
We also note that {\sc cloudy} models with {\sc bpass} v2.1 population synthesis
models calculated for a population of single stars or with 
{\sc starburst99} models predict a very low 
He~{\sc ii} $\lambda$4686 line intensity, $\la$ 0.1 per cent that of H$\beta$.

Adopting a top-heavy IMF, i.e. changing the IMF slope from the Salpeter value
$\alpha$ = $-$1.35 \citep{S55} to $\alpha$~=~$-$1,   and/or increasing the upper
mass limit from $M_{\rm up}$~=~100~M$_\odot$ to 300~M$_\odot$, will change the model
predictions in Fig.~\ref{fig3}a, but by not more than 10 per cent. Thus, we 
conclude that stellar radiation is unlikely to be the main source of hard 
ionizing photons for the production of high-ionization lines.

\subsection{Radiative shocks} \label{shock}

\citet{I12a} have suggested that interstellar radiative shocks with 
velocities of 300 -- 500 km s$^{-1}$ can produce sufficient extreme UV
ionizing radiation to reproduce observed 
He~{\sc ii} $\lambda$4686/H$\beta$ emission-line ratios of $\sim$ 2 -- 3 
per cent. 

To further investigate this possibility, we consider a set of 
spherically-symmetric composite {\sc cloudy} v17.01 models, 
adopting ionizing radiation consisting of
two components: a) radiation of a single stellar population with age 2 Myr, 
with a production rate of ionizing photons $Q_{\rm stellar}$ = 10$^{53}$ s$^{-1}$ 
corresponding to the average H$\beta$ luminosity of the two studied galaxies, and with various 
metallicities, calculated with the {\sc starburst99} code by using
Geneva evolutionary tracks of non-rotating stars; and b) radiation from 
radiative shocks coincident with the source of stellar radiation,
with various shock velocities and the lowest metallicity available, that
of the Small Magellanic Cloud, calculated by \citet{A08}. We set,  
somewhat arbitrarily, the production rate of ionizing photons from shocks 
$Q_{\rm shock}$ to be 5 -- 10 per cent 
of the production rate of ionizing stellar photons $Q_{\rm stellar}$. 
The ionization parameter $U$ averaged over the H~{\sc ii} region volume 
is varied in the 
range 10$^{-3.0}$ -- 10$^{-1.6}$, the same range as for the models discussed in 
Sect.~\ref{stellar}. Our models correspond to the case when shocks 
propagate through the ionized medium of the H~{\sc ii} region, different from 
the \citet{A08} calculations who considered propagation of shocks through
the neutral medium.

In Fig.~\ref{fig3}b we show the dependence of the 
He~{\sc ii} $\lambda$4686/H$\beta$ emission-line ratio on the shock velocity
for two composite models, with $Q_{\rm shock}$/$Q_{\rm stellar}$ = 5 per cent
(dashed line) and 10 per cent (solid line). The shaded region shows the range
of He~{\sc ii} $\lambda$4686/H$\beta$ for J0811$+$4730 and J1234$+$3901.
In contrast to the models with pure stellar ionizing radiation 
(Fig.~\ref{fig3}a), the composite model with 
$Q_{\rm shock}$/$Q_{\rm stellar}$ = 10 per cent can quite succesfully
reproduce observations for shock velocities of $\sim$~500~--~600~km~s$^{-1}$.

\citet{K18} discussing the origin of He~{\sc ii} $\lambda$4686 
emission line in SBS~0335$-$052E dismissed radiative shocks
as a source of this emission, citing the models by e.g. 
\citet{A08} which predict enhanced [O~{\sc i}] $\lambda$6300 and [S~{\sc ii}]
$\lambda$6717, 6731 emission lines compared to those observed in 
SBS~0335$-$052E. However, this argument is not convincing because 
\citet{A08} considered shocks propagating through a neutral ISM, while shocks in
SBS~0335$-$052E most likely propagate through an ionized medium. The H~{\sc ii} 
region extends to 1 -- 2 kpc from the ionizing clusters, and both the oxygen and 
sulfur in it are in stages of ionization higher than neutral oxygen and 
singly ionized sulfur.
Applying the composite models above to the SFGs J0811$+$4730 and J1234$+$3901, 
we find that radiative shocks increase at most the fluxes of 
the [O~{\sc i}] $\lambda$6300 and [S~{\sc ii}] $\lambda$6717, 6731 emission 
lines by $\sim$ 10 per cent compared to the models with pure stellar emission.

\begin{figure}
\centering{
\includegraphics[angle=-90,width=0.98\linewidth]{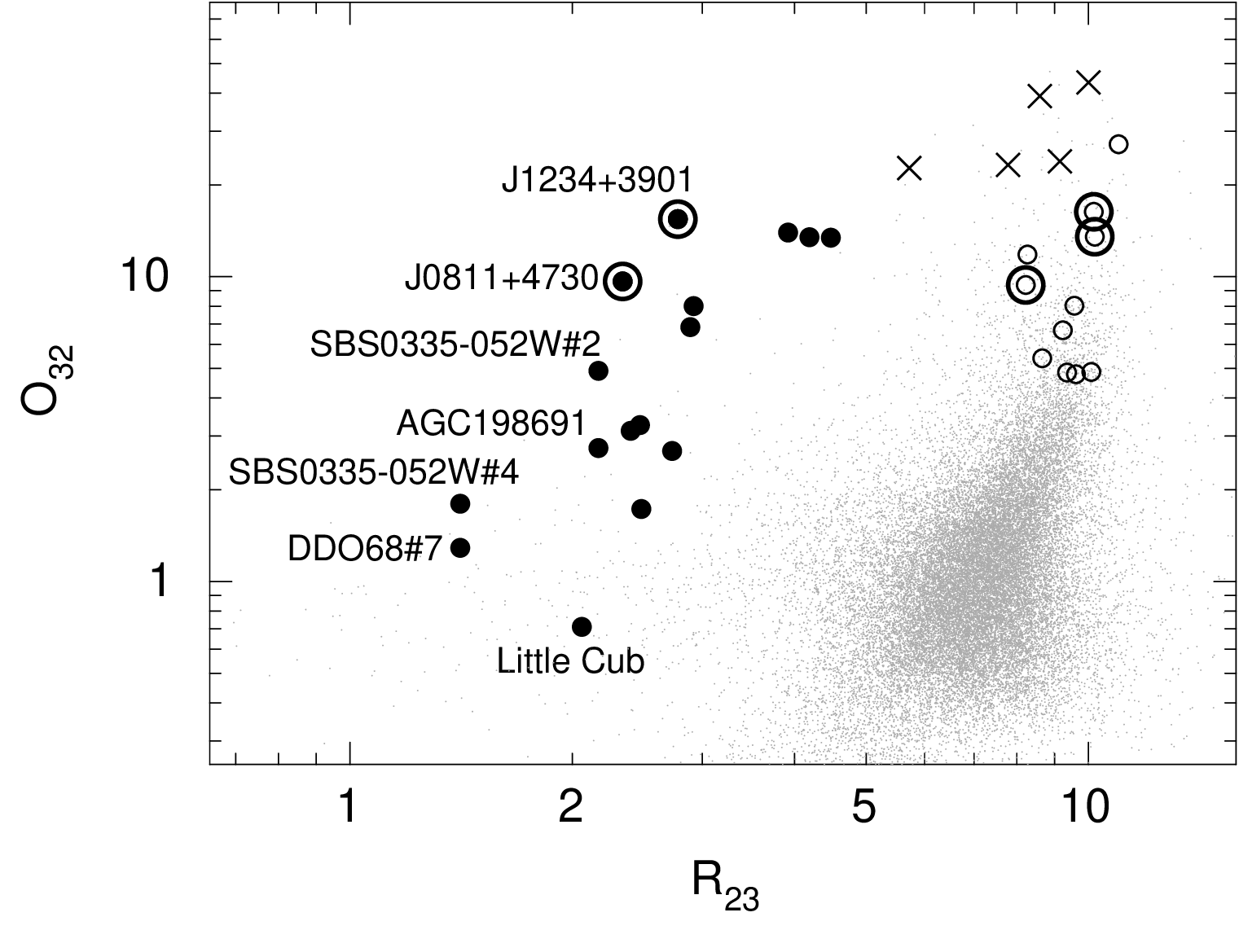}
}
\caption{The O$_{32}$ -- R$_{23}$ ratio for SFGs where O$_{32}$= 
[O~{\sc iii}]$\lambda$5007/[O~{\sc ii}]$\lambda$3727 and R$_{23}$ = 
([O~{\sc ii}]$\lambda$3727 + [O~{\sc iii}]$\lambda$4959 + 
[O~{\sc iii}]$\lambda$5007)/H$\beta$. The lowest-metallicity SFGs
from \citet[][A198691]{H16}, \citet[][Little Cub]{H17}, 
\citet[][DDO~68\#7]{An19} and \citet[][remaining galaxies]{I18a} are shown 
by filled circles and some of them are labelled.
J0811$+$4730 and J1234$+$3901 are shown by encircled filled circles. 
LyC leakers from 
\citet{I16a,I16b,I18b,I18c} are shown by open circles. The three objects 
with the highest LyC escape fractions are encircled. SFGs with the highest
O$_{32}$ are represented by crosses \citep{I17} and SFGs from the SDSS DR14 
by grey dots.}
\label{fig4}
\end{figure}

\section{Can J1234$+$3901 be a Lyman continuum leaking galaxy?}
\label{sec:diagrams}

It is generally thought that primeval galaxies were low-mass star-forming 
systems formed from
zero-metallicity gas at redshifts $\ga$ 10. They likely were the main sources 
of reionization of the Universe at redshifts $z$ $\sim$ 5 -- 10
\citep[e.g. ][]{O09,K17}. J1234$+$3901 as well as J0811$+$4730 are likely the 
best local counterparts of these primeval galaxies because of their 
compactness, low mass,
extremely low metallicity and vigorous ongoing star formation.

It has been suggested by \citet{JO13} and \citet{NO14} that one of the criteria 
for the selection of SFGs with density-bounded H~{\sc ii} regions, allowing 
Lyman continuum escape, is a high  O$_{32}$ = 
[O~{\sc iii}]$\lambda$5007/[O~{\sc ii}]$\lambda$3727 line ratio. 
Both J0811$+$4730 and J1234$+$3901 have very high O$_{32}$, respectively 
$\sim$10 and $\sim$15, implying LyC leakage. 

The O$_{32}$~--~R$_{23}$ diagram (where 
R$_{23}$ = 
([O~{\sc ii}]$\lambda$3727 + [O~{\sc iii}]$\lambda$4959 + 
[O~{\sc iii}]$\lambda$5007)/H$\beta$), including
both of the two lowest-metallicity galaxies discussed here, is shown in 
Fig.~\ref{fig4}. We have also plotted several other very metal-deficient local 
galaxies as listed in \citet{I18a} and one galaxy from \citet{H17}
(filled 
circles), SFGs with the highest O$_{32}$ known ($>$ 20, crosses) and 
$z$ $\sim$ 0.3 -- 0.4 LyC leakers (open circles and encircled open circles).
It is seen that J1234$+$3901 has the highest O$_{32}$ among the 
lowest-metallicity galaxies with 12 + logO/H $\la$ 7.3 (filled circles). 
Because of its extremely low metallicity, it is located very far to the left of
the SFGs with highest O$_{32}$ and the LyC leakers, as these have considerably 
higher 12 + logO/H $\sim$ 7.5 -- 8.0.
We note that both J0811$+$4730 and J1234$+$3901 have O$_{32}$ similar to those 
for LyC leakers with the highest escape fraction of ionizing radiation 
($f_{\rm esc}$(LyC) $\sim$ 46 -- 72 per cent) shown by encircled open circles 
\citep{I18b,I18c}. These similarities suggest that our 
two lowest-metallicity SFGs are likely also LyC leakers. 

However, \citet{I18c} have shown that the O$_{32}$ ratio is not a certain 
indicator of LyC leakage, because it depends also on other factors such as 
metallicity and
ionization parameter. Furthermore, because of their low redshifts, direct
observation of the ionizing radiation from these galaxies with the {\sl HST}/COS
is not possible. The most promising way to investigate whether J0811$+$4730 and
J1234$+$3901 are LyC leakers is to use an indirect indicator, the Ly$\alpha$
profile which can be observed with the {\sl HST}. Following the analysis of  
\citet{I17}, the presence of Ly$\alpha$ emission is expected in both galaxies 
from the values of their He~{\sc i} 3889/6678 and He~{\sc i} 7065/6678 
emission-line ratios, as derived from the LBT optical spectra which we have 
obtained for both BCDs. These ratios correspond to the case
of low optical depths $\tau$(3889) in the He~{\sc i} $\lambda$3889 emission
line and are similar to those in the confirmed
LyC leakers with strong Ly$\alpha$ emission \citep{I17}.

\section{Conclusions}\label{sec:conclusions}

In this paper we present Large Binocular Telescope (LBT)/Multi-Object Dual
Spectrograph (MODS) 
spectrophotometric observations of the star-forming galaxy (SFG)
J1234$+$3901, selected from the Data Release 14 (DR14) of the Sloan Digital Sky
Survey (SDSS). This galaxy can also be classified as a blue compact dwarf (BCD)
galaxy because of its low mass, blue colour indicating active star formation
and compact structure. Our main results are as follows.

1. The properties of J1234$+$3901 are extreme in many ways. Its oxygen abundance
is 12+logO/H = 7.035$\pm$0.026, 
one of the lowest for nearby SFGs. This galaxy with a redshift
$z$ = 0.133 is the most distant and most luminous object known among BCDs with 
12 + logO/H $\la$ 7.3. Its mass-to-luminosity ratio of $\sim$ 0.01 in solar
units is extremely low, indicating that its optical luminosity is strongly
dominated by a very young stellar population. These properties are very 
similar to those of the most metal-deficient SFG known, J0811$+$4730 
\citep{I18a}. Both galaxies deviate strongly from the
luminosity-metallicity and stellar mass-metallicity relations defined by the 
bulk of SFGs: they are 3-5 more metal-poor for their stellar masses and 
about a factor of 5-8 more metal-deficient for their SDSS $g$-band luminosities.

2.  A strong He~{\sc ii} $\lambda$4686 emission line with a flux of 2.4 per cent
that of H$\beta$ is observed in the spectrum of J1234$+$3901. We discuss
possible sources of hard ionizing radiation that can produce He~{\sc ii} 
emission and conclude that stellar ionizing radiation is insufficient.
Comparing model predictions and the observed fluxes of high-ionization lines 
in  J1234$+$3901 and the most metal-deficient SFG known J0811$+$4730, 
we find that the most likely source of hard ionizing radiation in these 
galaxies is fast radiative shocks. 

3. The O$_{32}$ = [O~{\sc iii}]$\lambda$5007/[O~{\sc ii}]$\lambda$3727 flux
ratios of $\sim$ 10 and of $\sim$ 15 in J0811$+$4730 and J1234$+$3901,
respectively, are very high. The O$_{32}$ value for J1234$+$3901 is the highest 
among the lowest-metallicity SFGs. These properties imply that the two galaxies 
are likely Lyman continuum leakers and that they are
the best local counterparts of high-redshift primeval SFGs.

\section*{Acknowledgements}

Y.I.I. and N.G.G. acknowledge support from the National Academy of Sciences
of Ukraine (Project No. 0116U003191) and by its Program of Fundamental
Research of the Department of Physics and Astronomy (Project No. 0117U000240).
Funding for the Sloan Digital Sky Survey IV has been provided by
the Alfred P. Sloan Foundation, the U.S. Department of Energy Office of
Science, and the Participating Institutions. SDSS-IV acknowledges
support and resources from the Center for High-Performance Computing at
the University of Utah. The SDSS web site is www.sdss.org.
SDSS-IV is managed by the Astrophysical Research Consortium for the 
Participating Institutions of the SDSS Collaboration. 
This research has made use of the NASA/IPAC Extragalactic Database (NED), which 
is operated by the Jet Propulsion Laboratory, California Institute of 
Technology, under contract with the National Aeronautics and Space 
Administration.

\bsp

\label{lastpage}


\begin{thebibliography}{}



\bibitem[Abolfathi et al.(2018)]{A18} Abolfathi B. et al., 2018, \apjs, 235, 42


\bibitem[Allen et al.(2008)]{A08} Allen M. G., Groves B. A., Dopita M. A., 
Sutherland R. S., Kewley L. J., 2008, \apjs, 178, 20

\bibitem[Aller(1984)]{A84} Aller L. H., 1984, Physics of Thermal
Gaseous Nebulae (Dordrecht: Reidel)




\bibitem[Annibali et al.(2019)]{An19} Annibali F. et al., 2019, \mnras, 
482, 3892








\bibitem[Blanton et al.(2003)]{B03} Blanton M. R. et al., 2003, \apj, 592, 819 



















\bibitem[Eldridge et al.(2017)]{E17} Eldridge J. J., Stanway E. R., Xiao L., 
McClelland L. A. S., Taylor G., Ng M., Greis S. M. L., Bray J. C., 
2017, \pasa, 34, 58




\bibitem[Ferland et al.(2017)]{F17} Ferland G. J. et al., 2017, \rmxaa, 
53, 385

\bibitem[Filippenko(1982)]{F82} Filippenko A. V., 1982, \pasp, 
94, 715








\bibitem[Girardi et al.(2000)]{G00} Girardi L., Bressan A., Bertelli G., 
Chiosi C., 2000, \aaps, 141, 371



\bibitem[Guseva et al.(2000)Guseva, Izotov \& Thuan]{GIT00} Guseva N. G., 
Izotov Y. I., Thuan T. X., 2000, \apj, 531, 776


\bibitem[Guseva et al.(2017)]{G17} Guseva N. G., Izotov Y. I., Fricke K. J.,
Henkel C., 2017, \aap, 599, A65


\bibitem[Hirschauer et al.(2016)]{H16} Hirschauer A. S. et al., 2016, \apj, 822,
108


\bibitem[Hsyu et al.(2017)]{H17} Hsyu, T., Cooke, R. J., Prochaska, J. X., 
Bolte, M. 2017, \apj, 845, L22

\bibitem[Hunt et al.(2015)]{H15} Hunt L. K. et al., 2015, \aap, 583, A114


\bibitem[Izotov \& Thuan(1998)]{IT98} Izotov Y. I., Thuan T. X., 
1998, \apj, 497, 227



\bibitem[Izotov et al.(1994)Izotov, Thuan \& Lipovetsky]{ITL94} Izotov Y. I.,
Thuan T. X., Lipovetsky V. A., 1994, \apj, 435, 647

\bibitem[Izotov et al.(2001)Izotov, Chaffee \& Schaerer]{I01} Izotov Y. I., 
Chaffee F. H., Schaerer D., 2001, \aap, 378, L45

\bibitem[Izotov et al.(2004)]{I04} Izotov Y. I., Noeske K. G., Guseva N. G., 
Papaderos P., Thuan T. X., Fricke K. J., 2004, \aap, 415, L27

\bibitem[Izotov et al.(2005)Izotov, Thuan \& Guseva]{I05} Izotov Y. I., 
Thuan T. X., Guseva N. G., 2005, \apj, 632, 210

\bibitem[Izotov et al.(2006a)]{I06a} Izotov Y. I., Stasi\'nska G., Meynet G., 
Guseva N. G., Thuan T. X., 2006a, \aap, 448, 955

\bibitem[Izotov et al.(2006b)]{I06b} Izotov Y. I., Schaerer D., Blecha A., 
Royer F., Guseva N. G., North P., 2006b, \aap, 459, 71



\bibitem[Izotov et al.(2009)]{I09} Izotov Y. I., Guseva N. G., 
Fricke K. J., Papaderos P., 2009, \aap, 503, 61


\bibitem[Izotov et al.(2012a)Izotov, Thuan \& Guseva]{I12a} Izotov Y. I., 
Thuan T. X., Guseva N. G., 2012a, \aap, 546, 122

\bibitem[Izotov et al.(2012b)Izotov, Thuan \& Privon]{I12b} Izotov Y. I., 
Thuan T. X., Privon G., 2012b, \mnras, 427, 1229


\bibitem[Izotov et al.(2014)]{I14} Izotov Y. I., Guseva N. G., 
Fricke K. J., Henkel C., 2014, \aap, 561, A33







\bibitem[Izotov et al.(2016a)]{I16a} Izotov Y. I., Orlitov\'a I., Schaerer D., 
Thuan T. X., Verhamme A., Guseva N. G., Worseck G., 2016a, \nat, 529, 178

\bibitem[Izotov et al.(2016b)]{I16b} Izotov Y. I., Schaerer D., Thuan, T. X., 
Worseck G., Guseva N. G., Orlitov\'a I., Verhamme A., 2016b, \mnras, 461, 3683

\bibitem[Izotov et al.(2017)Izotov, Thuan \& Guseva]{I17} Izotov Y. I., 
Thuan T. X., Guseva N. G., 2017, \mnras, 471, 548

\bibitem[Izotov et al.(2018a)]{I18a} Izotov Y. I., Thuan T. X., Guseva N. G.,
Liss S. E., 2018a, \mnras, 473, 1956

\bibitem[Izotov et al.(2018b)]{I18b} Izotov Y. I., Schaerer D., Worseck G., 
Guseva N. G., Thuan T. X., Verhamme A., Orlitov\'a I., Fricke K. J.,
2018b, \mnras, 474, 4514

\bibitem[Izotov et al.(2018c)]{I18c} Izotov Y. I., Worseck G., Schaerer D., 
Guseva N. G., Thuan T. X., Fricke K. J., Verhamme A., Orlitov\'a I., 
2018c, \mnras, 478, 4851

\bibitem[Jaskot \& Oey(2013)]{JO13} Jaskot A. E., Oey M. S.,
2013, \apj, 766, 91

\bibitem[Karman et al.(2017)]{K17} Karman W. et al., 2016, \aap, 599, A28



\bibitem[Kehrig et al.(2018)]{K18} Kehrig C., V\'ilchez J. M., Guerrero M. A., 
Iglesias-P\'aramo J., Hunt L. K., Duarte-Puertas S., Ramos-Larios G.,
2018, \mnras, 480, 1081

\bibitem[Kennicutt(1998)]{K98} Kennicutt R. C., Jr.,
1998, \araa, 36, 189








\bibitem[Leitherer et al.(1999)]{L99} Leitherer C. et al., 1999, \apjs, 123, 3

\bibitem[Leitherer et al.(2014)]{L14} Leitherer C., Ekstr\"om S., 
Meynet G., Schaerer D., Agienko K. B., Levesque E. M., 2014, \apjs, 212, 14


\bibitem[Lejeune et al.(1997)Lejeune, Buser \& Cuisiner]{L97} 
Lejeune T., Buser R., Cuisinier F., 1997, \aaps, 125, 229







\bibitem[Meynet et al.(1994)]{M94} Meynet G., Maeder A., Schaller G., 
Schaerer D., Charbonnel C., 1994, \aaps, 103, 97




\bibitem[Nakajima \& Ouchi(2014)]{NO14} Nakajima K., Ouchi M., 2014, 
\mnras, 442, 900



\bibitem[Ouchi et al.(2009)]{O09} Ouchi M. et al., 2009, \apj, 706, 1136





\bibitem[Planck Collaboration XVI(2014)]{P14} Planck Collaboration XVI,
2014, \aap, 571, A16















\bibitem[Salpeter(1955)]{S55} Salpeter E. E., 1955, \apj, 121, 161




\bibitem[Searle \& Sargent(1972)]{SS72} Searle L., Sargent W. L. W., 1972,
\apj, 173, 25

\bibitem[Shirazi \& Brinchmann(2012)]{SB12} Shirazi M., Brinchmann J.,
2012, \mnras, 421, 1043


\bibitem[Skillman \& Kennicutt(1993)]{SK93} Skillman E., Kennicutt R. C. Jr.,
 1993, \apj, 411, 655


\bibitem[Smith et al.(2002)Smith, Norris \& Crowther]{S02} 
Smith L. J., Norris R. P. F., Crowther P. A., 2002, \mnras, 337, 1309


\bibitem[Stasi\'nska \& Izotov(2003)]{SI03} Stasi\'nska G., Izotov Y., 
2003, \aap, 397, 71





\bibitem[Thuan \& Martin(1981)]{TM81} Thuan T. X., Martin G. E., 1981, \apj,
\apj, 247, 823

\bibitem[Thuan \& Izotov(2005)]{TI05} Thuan T. X., Izotov Y. I., 2005, \apj,
\apjs, 161, 240



















\end{thebibliography}
\end{document}